\documentclass[aps,showpacs, nofootinbib]{revtex4}

\newcommand{\be}{\begin{equation}}
\newcommand{\ee}{\end{equation}}
\newcommand{\ben}{\begin{eqnarray}}
\newcommand{\een}{\end{eqnarray}}

\newcommand{\la}{{\lambda}}

\newcommand{\cO}{{\cal O}}

\newcommand{\cD}{{\cal D}}
\newcommand{\cM}{{\cal M}}
\newcommand{\cJ}{{\cal J}}

\newcommand{\p}{\partial}
\newcommand{\na}{\nabla}

\newcommand{\tA}{\tilde A}

\newcommand{\ep}{\epsilon}
\newcommand{\bep}{\bar \epsilon}

\newcommand{\ga}{\gamma}

\pacs{04.70.Bw, 04.50.Kd, 04.20.-q}

\begin{document}

\title{Uniqueness of dilaton Melvin-Schwarzschild solution}

\author{Marek Rogatko}
\affiliation{Institute of Physics \protect \\
Maria Curie-Sklodowska University \protect \\
20-031 Lublin, pl.~Marii Curie-Sklodowskiej 1, Poland \protect \\
marek.rogatko@poczta.umcs.lublin.pl \protect \\
rogat@kft.umcs.lublin.pl}

\date{\today}

\begin{abstract}
We show that dilaton Melvin-Schwarzschild black hole solution is the only
asymptotically dilaton Melvin static axisymmetric black hole solution of Einstein-Maxwell dilaton
equations of motion.

\end{abstract}

\maketitle

\section{Introduction}
Melvin solution in Einstein-Maxwell gravity describes a regular and static cylindrically
symmetric solution representing a bundle of magnetic flux lines in magnetostatic-gravitational
equilibrium \cite{mel64,mel65}. On the other hand, it provides a uniform magnetic field in the realm
of general relativity. Next, 
the considerations pertaining to the description of 
a rotating time-dependent magnetic universe were conceived in \cite{gar94}, while the 
problem of gravitational waves traveling through the universe in question
was tackled in Ref.\cite{gar92}.
\par
There was also a great interest in black holes behavior in magnetic universe. For the first time,
the exact solution
describing a black hole in Melvin universe was achieved in \cite{ern76}.  The ultrastatic boost of 
Schwarzschild black hole immersed in an external electromagnetic 
field was studied in \cite{ort04}, while magnetized static black hole acting as a gravitational lens was 
examined in Ref.\cite{kon07}.
Detailed studies of ergoregions and thermodynamics of magnetized black hole solutions were conducted in 
Refs.\cite{gib13} (see also the earlier works 
connected with black holes in magnetized universe,
e.g., \cite{ear}), while the studies of spectral line broadening effect in spacetime of non-rotating magnetized black hole
were elaborated in \cite{fro14}. The problem of magnetic fields in the expanding universe was tackled in \cite{kas14}.
Magnetized black hole solutions were also established in generalization of Einstein-Maxwell theory, in heterotic 
string theory, were
a Melvin universe with nontrivial dilaton and axion fields was founded \cite{tse95,gib88}. On the other hand, a 
Melvin flux tube universe can be also realized as the Kaluza-Klein reduction from a flat higher
spacetime with nontrivial identifications. Such
solution generating technique which allows to embed
the dilaton C-metric in dilaton magnetic universe was presented \cite{dow94a}, whereas the case of a pair creation of the extremal black hole and Kaluza-Klein monopoles
was examined in \cite{dow94b}. Melvin type solution was also studied in gravity theories minimally coupled to any nonlinear electromagnetic theory, including 
Born-Infeld electrodynamics \cite{gib01}.
\par
A new Melvin-like solution with a Liouville type potential was given in Ref.\cite{rad04}, whereas electrically charged dilaton black hole in magnetic field was considered in \cite{yaz13}.
The generalization of the aforementioned problems in higher dimensional gravity was given in \cite{ort05}, where magnetized static and rotating black holes in arbitrary dimensions and
magnetized black rings in five-dimensions were examined. The researches were based on applying a Harrison transformation to the known black hole and black ring solutions.
On the other hand, the higher dimensional dilaton gravity case was treated in Ref.\cite{yaz06}.
This subject is also treated in M-theory, where a version of Melvin universe
was found in type II A and type 0A string theory \cite{rus98}.\\
The subject of black hole magnetized solutions was also treated by Ernst's solution generating technique, where black hole solutions in magnetic universe in Einstein-Maxwell theory coupled 
conformally to a scalar field \cite{ast13}, as well as, axisymmetric stationary black holes with cosmological constant \cite{ast12} were considered. These 
techniques were used to obtain a C-metric with conformally coupled scalar field in magnetic universe \cite{ast13b}
and exact and regular solution describing a couple of charged spinning black hole in an external electromagnetic field \cite{ast14}.
Recently, there was presented a solution of Einstein-Maxwell theory unifying both magnetic Bertotti-Robinson and Melvin solution as a single axisymmetric line element \cite{maz13}.
\par
On the other hand,
gravitational collapse and emergence of black holes were paid attention to. The most
intriguing was Wheller's {\it black hole no-hair} conjecture or its mathematical formulation, the uniqueness theorem 
(problem of classification of domains of outer communication of suitably regular black hole spacetimes). The first 
attempts to classify non-singular static black hole solutions in Einstein gravity was undertaken in
\cite{isr}
and some more results were gained in Refs.\cite{mil73}-\cite{he93}.
The complete classification of static vacuum and electro-vacuum black hole solutions was finished 
in \cite{chr99a,chr99b}, where the condition of non-degeneracy of the 
event horizon was removed  as well as it was proved that all degenerate components of 
the black hole event horizon have charges of the same signs.
\par
As far as stationary axisymmetric black holes is concerned, the problem turned out to be far more complicated
\cite{stat} and the complete uniqueness proof was achieved by Mazur \cite{maz} and Bunting \cite{bun}
(see for a review of the uniqueness of black hole
solutions story see \cite{book} and references therein).
In higher dimensional generalization of gravity theory motivated by the contemporary unifications schemes 
such as M/string theories the classification of higher dimensional 
charged black holes both with non-degenerate and degenerate component 
of the event horizon was proposed in Refs.\cite{nd}, while some progress concerning the nontrivial case 
of $n$-dimensional rotating black objects (black holes, black rings or black lenses) uniqueness theorem were 
presented in \cite{nrot}. On the other hand, the behavior of matter fields in the spacetime of higher dimensional 
black hole was examined in \cite{rog12}.
\par
The desire of constructing a consistent quantum gravity theory triggered also interests 
in mathematical aspects of black holes in the low-energy limit of the string theories
and supergravity \cite{sugra}. On the other hand, various modifications of Einstein gravity 
such as the Gauss-Bonnet extension were examined from the point of view of black hole uniqueness theorem.
The strictly stationary static vacuum spacetimes was discussed in \cite{shi13a}, while it turned out that
up to the small curvature limit, static uncharged or electrically charged Gauss-Bonnet black hole is diffeomorphic to 
Schwarzschild-Tangherlini or Reissner-Nordstr\"om black hole solution, respectively \cite{rog14}. On the other hand, 
in Chern-Simons modified gravity it was proved that a static asymptotically flat black hole
solution is unique to be Schwarzschild spacetime \cite{shi13}, while electrically 
charged black hole in the theory in question is diffeomorphic to
Reissner-Nordstr\"om black hole \cite{rog13}.
\par
Motivated by the aforementioned problems 
concerning both magnetized axially symmetric black hole solutions as well as their classification,
we shall consider the problem of the uniqueness static dilaton magnetized Schwarzschild black hole. 
The first attempts to establish the uniqueness of the static Melvin solution in Einstein-Maxwell
gravity was attributed to Hiscock \cite{his81}, where the generalization of the Israel's theorem was presented.
In our research, we prove that
dilaton Melvin-Schwarzschild black hole solution is the only
asymptotically dilaton Melvin static axisymmetric black hole solution of Einstein-Maxwell dilaton
equations of motion. 
\par
The organization of our paper is as follows. Sec.I is devoted to the equations of motion for a static 
axially symmetric system with $U(1)$-gauge Maxwell matter source, in the low-energy
limit of the heterotic string theory. In Sec.II we 
provide the uniqueness theorem proof revealing that the only static axially symmetric solution in the theory in question
with $S^2$-topology of the event horizon is the Melvin-Schwarzschild dilaton black hole.
In Sec.III we conclude our investigations.

\section{Equations of motion}
Our main aim of the paper will be to show the uniqueness of static axisymetric magnetized black hole,
the so-called dilaton Melvin-Schwarzschild black hole,
in dilaton gravity being the low-energy limit of the heterotic string theory. In this section we
analyze the equation of motion for static axisymmetric line element in the theory under consideration,
which action yields 
\be
I = \int d^4 x \sqrt{-g} 
\bigg( R - 2(\na \varphi)^{2}
- e^{-2\varphi} F_{\alpha \beta} F^{\alpha \beta} \bigg),
\label{act}
\ee
where the strength of the Maxwell gauge field is described by
$F_{\mu \nu} = 2\na_{[\mu} A_{\nu]}$.
The resulting equations of motion, derived from the variational
principle, are provided by
\ben
R_{\mu \nu} = e^{-2 \varphi} \bigg(
2 F_{\mu \rho} F_{\nu}{}{}^{\rho} - {1 \over 2} g_{\mu \nu}F^2 \bigg) +
2 \na_{\mu} \varphi \na_{\nu} \varphi, \\
\na_{\mu} \na^{\mu} \varphi + {1 \over 2} e^{-2\varphi} F^2 = 0,\\ 
\na_{\mu} \left ( e^{-2 \varphi} F^{\mu \nu} \right ) = 0.
\een
A line element appropriate to the static axisymmetric spacetime can
be expressed as
\be
ds^2 = - e^{2 \psi} dt^2 + e^{-2 \psi} \left [
e^{2 \ga} \left ( d\rho^2 + dz^2 \right ) + \rho^2 d\phi^2 \right ],
\label{gij}
\ee
where the functions $\psi $ and $\ga$ depended only on $\rho$ and $z$ coordinates.
In our considerations we assume that the non-zero components of the $U(1)$-gauge strength tensor
will be in the form $A_{\mu } = (A_{t},~ 0,~ 0, A_\phi)$.
Moreover, one supposes that the gauge field components
are also functions of $\rho$ and $ z$. Consequently,
the underlying equations of motion imply
\be
\na^2 \varphi - e^{-2\psi - 2\varphi} \bigg(
A_{t, \rho}^2 + A_{t, z}^2 \bigg) + {e^{2\psi - 2 \varphi} \over \rho^2}
\bigg( A_{\phi, \rho}^2 + A_{\phi, z}^2 \bigg) = 0,
\label{em1}
\ee
\be
\na^2 \psi - e^{-2\psi - 2\varphi} \bigg(
A_{t, \rho}^2 + A_{t, z}^2 \bigg) - {e^{2\psi - 2 \varphi} \over \rho^2}
\bigg( A_{\phi, \rho}^2 + A_{\phi, z}^2 \bigg) = 0,
\label{em2}
\ee
\be
\na^2 A_{t} - 2 \bigg( \psi_{, \rho} + \varphi_{, \rho} \bigg) A_{t, \rho}
- 2 \bigg( \psi_{, z} + \varphi_{, z} \bigg) A_{t, z} = 0,
\label{em3}
\ee
\be
\triangle A_{\phi} +
2 \bigg( \psi_{, \rho} - \varphi_{, \rho} \bigg) A_{\phi, \rho} +
2 \bigg( \psi_{, z} - \varphi_{, z} \bigg) A_{\phi, z} = 0,
\label{em4}
\ee
\be
e^{-2\psi - 2\varphi} \bigg(
A_{t, \rho}^2 - A_{t, z}^2 \bigg) + {1 \over \rho^2} e^{2\psi - 2 \varphi} 
\bigg( A_{\phi, \rho}^2 - A_{\phi, z}^2 \bigg) +
\bigg( \varphi_{,z}^2 - \varphi_{, \rho}^2 \bigg) =
\psi_{, \rho}^2 - \psi_{, z}^2 - {\ga_{, \rho} \over \rho},
\label{em5}
\ee
\be
{\ga_{, z} \over \rho} - 2 \psi_{, \rho} \psi_{, z} =
- 2 e^{-2\psi - 2\varphi} A_{t, \rho} A_{t, z} + {2 \over \rho^2} e^{2\psi - 2 \varphi} 
A_{\phi, \rho} A_{\phi, z} + 2 \varphi_{, \rho} \varphi_{, z},
\label{em6}
\ee
where $\na^2$ is the Laplacian operator in the $(\rho,~ z)$ coordinates,
namely, $\na^2 = \p_{\rho}^2 + \p_{z}^2 + {1 \over \rho} \p_{\rho}$
and $\triangle = \p_{\rho}^2 + \p_{z}^2 - {1 \over \rho}\p_{\rho}$.\\
From equation (\ref{em5}) or (\ref{em6}) one can determine the function $\ga$
if $\psi,~ \varphi,~ A_{t},~ A_{\phi}$ are known. 
By virtue of this, for our present purposes it will be sufficient to consider
the relations given by (\ref{em1}-\ref{em4}).
\par
To proceed further, one defines the quantities provided by the following relations:
\be 
K = - \varphi - \psi, \qquad L = \psi - \varphi.
\ee
They enable us to rewrite the above system of partial differential equations
(\ref{em1})-(\ref{em4}), in the forms as
\be
\na^2 K + e^{2 K} \na \tA_{t}~ \na \tA_{t} = 0,
\label{a0}
\ee
\be
\na^2 \tA_{t} + 2 \na K ~ \na \tA_{t} = 0,
\label{a1}
\ee
\be
\na^2 L + e^{2 L} \na \tA_{\phi}~  \na \tA_{\phi} = 0,
\label{a2}
\ee
\be
\triangle \tA_{\phi} + 2 \na L ~\na \tA_{\phi} = 0,
\label{a3}
\ee
where we have defined the quantities which imply
\be
\tA_{t} = {A_{t} \over \sqrt{2}}, \qquad 
\tA_{\phi} = {i ~A_{\phi} \over \sqrt{2}}.
\ee
As a result, we have obtained two pairs of equations described in terms of 
$\tA_{t},~ K$ and the other for $\tA_{\phi}$ and $L$.\\
On the other hand, one notices that equation (\ref{a1}) allows us to specify a pseudopotential
$\eta$ which yields
\be
\eta_{\rho} = \rho~ e^{2 K} \tA_{t, z}, \qquad
\eta_{z} = - \rho~ e^{2 K} \tA_{t, \rho},
\label{wa}
\ee
while the relation (\ref{a3}) leads to the following
pseudopotential $\chi$:
\be
\chi_{\rho} =  - {e^{2 L} \tA_{\phi, z} \over \rho}, \qquad
\chi_{z} = {e^{2 L} \tA_{\phi, \rho} \over \rho}.
\label{wb}
\ee
Our main aim is to rewrite the system of equations (\ref{a0}-\ref{a3}) 
in the forms similar to the Ernst ones.
In order
to do this, we introduce two complex scalars, determined by
\ben
\label{e1}
\ep_{1} &=& \rho~ e^K + i~ \eta, \\
\ep_{2} &=&  e^{L} + i~ \chi.
\een
The complex scalar allow us to arrange equations
(\ref{a0}-\ref{a1}) and (\ref{a2}-\ref{a3}) in the following
two complex relations:
\ben
\label{ee1}
\bigg( \bep_{1} + \ep_{1} \bigg)~ \na^2 \ep_{1} =
2~ \na \ep_{1} \na \ep_{1}, \\ \label{ee2}
\bigg( \bep_{2} + \ep_{2} \bigg)~ \na^2 \ep_{2} =
2~ \na \ep_{2} \na \ep_{2},
\een
where a bar denotes complex conjugation.\\
The received relations (\ref{ee1}) and (\ref{ee2}) enable us to
combine in a convenient and a symmetric fashion the two equations governing
$\varphi,~ \psi$ and the adequate components of $U(1)$-gauge field.
Each of the discussed relation constitute a sigma model.

\subsection{Boundary conditions}
In this subsection we pay attention to the relevant boundary conditions in the case
under consideration. As was pointed out,
dilaton Melvin-Schwarzschild black hole \cite{yaz06} obtained by the implementation of a Harrison
transformation to the known black hole solution,
asymptotically approaches the so-called dilaton Melvin solution
given in Ref.\cite{dow94a}. In order to provide some continuity with the known results concerning
stationary axisymmetric as well as Einstein-Maxwell Melvin black hole uniqueness theorems, we introduce
in two-dimensional manifold the spheroidal coordinates provided by
\be
\rho^2 = (\la^2 - c^2)~(1 - \mu^2), \qquad z = \la~\mu,
\ee
where $\mu = \cos \theta$ is chosen in such a way that the black hole event horizon boundary is situated 
at a constant value of $\la = c$.
On the other hand, two rotation axis segments which distinguish the south and the north segments of the event horizon
are described by the respective limit $\mu = \pm 1$.
Just we obtain the line element in the form as
\be
d\rho^2 + dz^2 = (\la^2 - \mu^2~c^2)\bigg(
{d\la^2 \over \la^2 - c^2} + {d\mu^2 \over 1 - \mu^2} \bigg).
\ee
We also introduce the quantities which yield
\be
X= e^{-2 \psi}~\rho^2, \qquad e^{- 2 \psi + 2 \ga} = {e^h \over X},
\ee
where $h$ is a function of $(\rho,~z)$-coordinates. 
Introducing such coordinates we assume that the black hole event horizon has topology of
$S^2$ sphere.
\par
In what follows one will examine the 
domain of outer communication $<<\cD>>$ as a rectangle. Namely, one gets
\ben
\p \cD^{(1)} &=& \{ \mu = 1,~\la= c, \dots, R \},\\ \nonumber
\p \cD^{(2)} &=& \{ \la = c,~\mu = 1, \dots, -1 \},\\ \nonumber
\p \cD^{(3)} &=& \{ \mu = - 1,~\la= c, \dots, R \},\\ \nonumber
\p \cD^{(4)} &=& \{ \la  = R,~\mu= -1, \dots, 1 \}.
\een
The relevant boundary conditions may be cast as follows. 
At infinity, we insist that
$X,~A_\phi, ~A_t$ are well-behaved functions and the solution in question asymptotically tends
to the Melvin-dilaton one. It implies the following behaviors:
\ben \label{inf}
{1 \over \rho^2}~X &=& \bigg( 1 + {1 \over 2}~B^2~\rho^2 \bigg)^{-2}~\bigg( 1 + \cO(\la^{-1})
\bigg),\\
A_\phi &=& {B~\rho^2 \over 2~\bigg( 1 + {B^2~\rho^2 \over 2} \bigg)}~
\bigg( 1 + \cO(\la^{-1})\bigg),\\
A_t &=& \cO(\la^{-1}), \qquad \varphi = \cO(\la^{-1}),
\een
where now $\rho$ stands for the asymptotical cylindrical coordinate given by
$\rho^2 \rightarrow \la^2~(1 - \mu^2)$. 
On the black hole event horizon,
where $\la \rightarrow c$, the quantities in question behave regularly (see, e.g., \cite{stat,maz,book}).
Namely, they yield
\ben \label{hor}
X &=& \cO(1 - \mu^2), \qquad {1 \over X}~\p_\mu~X = - {2 \mu \over 1 - \mu^2} + \cO(1),\\
\p_\la~A_\phi &=& \cO(1 - \mu^2), \qquad \p_\mu~A_\phi = \cO(1), \\
\p_\la~A_t &=& \cO(1), \qquad  \p_\mu~A_t = \cO(1), \qquad \varphi = \cO(1).
\een
On the other hand, on the symmetry axis as $\mu \rightarrow \pm 1$,
we obtain
\ben \label{ax}
X &=& \cO(1), \qquad {1 \over X} = \cO(1),\\
A_\phi &=& \cO(1), \qquad \p_\la~A_\phi = \cO(1), \qquad \p_\la \varphi = \cO(1),\\
A_t &=& \cO(1), \qquad \p_\la A_t = \cO(1), \qquad \varphi = \cO(1).
\een

\section{Uniqueness of Black Hole Solutions}
In order to proceed to the uniqueness proof, we recall that \cite{gur82,gur84} Ernst's equations can be included
in the single matrix equation. The matrix equation written out in terms of the adequate elements represents relation
which constitute various combinations of the aforementioned equations. In the case under consideration
the matrix equation can be expressed as
\be
\p_{\rho} \left [ P_{(i)}^{-1} \p_{\rho} P_{(i)} \right ]
+ \p_{z} \left [ P_{(i)}^{-1} \p_{z} P_{(i)} \right ] = 0,
\label{ma}
\ee
where the subscript $(i)$ in $P$ matrix 
refers respectively to $\tA_t,~ \tA_\phi$ gauge
fields. 
The explicit form  of the matrices are given by
\be
P_{(t)} = {1 \over \rho~ e^K} \pmatrix{1 & \eta \cr \eta &
\rho^2~ e^{2 K} + \eta^2 \cr},
\qquad
P_{(\phi)} = {1 \over e^L} \pmatrix{1 & \chi \cr \chi &
e^{2 L} + \chi^2 \cr}.
\ee
By simple calculations one can check that, the above matrix equations when written
out explicitly in terms of its elements constitute four relations, all of
them are various combinations of the equations (\ref{ee1}-\ref{ee2}) as well as their
complex conjugations. 
It can be noticed that when $B$ is any constant invertible matrix, then $B~P~B^{-1}$ represents
also solution of the matrix equation in question. In principle, choosing different forms
of the matrix $B$ one is able to derive all the transformations which are applicable to the system of
Ernst's equations.
\par
In order to prove a uniqueness theorem we shall follow the line
described in Refs.\cite{maz,gur}. 
To begin with, one should assume enough
differentiability for the matrices components in a domain of outer communication  $<<\cD>>$
of the two-dimensional manifold $\cM$ with boundary $\p \cD$.
Let $P_{(i) 1}$ and $P_{(i) 2}$ will be two different solutions of the equation (\ref{ma}),
respectively for the cases of $\tA_0$ and $\tA_\phi$ components of $U(1)$-gauge field. The difference
of the above relations can be written as
\be
\na \left (  P_{(i) 1}^{-1} \left ( \na Q_{(i)} \right )
P_{(i) 2} \right ) = 0, 
\label{dif}
\ee
where $Q_{(i)} = P_{(i) 1} P_{(i) 2}^{-1}$ and $i = t,~\phi$.
Multiplying the above relations by $Q_{(i)}^{\dagger}$ and taking the trace, we get the expression which 
implies
\be
\na^2 q_{(i)} = Tr \left [
\left ( \na Q_{(i)}^{\dagger} \right ) P_{(i) 1}^{-1}
\left ( \na Q_{(i)} \right ) P_{(i) 2} \right ],
\label{g}
\ee
where we have denoted by $q_{(i)} = Tr Q_{(i)}$. Having in mind hermicity and
positive definiteness of the matrices $P_{(i) 1}$ and $P_{(i) 2}$,
we can postulate the following form of the above matrices:
\be
P_{(t)} = M ~M^{\dagger},
\qquad
P_{(\phi)} = N~ N^{\dagger}.
\label{pa}
\ee
It can be revealed that the explicit forms of the matrices
$M$ and $N$ are provided by 
\be
M = {1 \over \sqrt{\rho}~ e^{{K \over 2}}} \pmatrix{
1 & 0 \cr \eta & \rho~ e^{K} \cr},
\qquad
N_{\alpha} = {1 \over e^{L \over 2}} \pmatrix{
1 & 0 \cr \chi & e^{L} \cr}.
\label{aa}
\ee
Relation (\ref{pa}) enables one to rewrite equation (\ref{g}) in the form which yields
\be
\na^2 q_{(i)} = Tr \left ( \cJ_{(i)}^{\dagger} \cJ_{(i)} \right ),
\label{jj}
\ee
where $ \cJ_{(t)} = M_{1}^{-1} (\na Q_{(A)}) M_{2}$ and
$\cJ_{(\phi)} = N_{1}^{-1} (\na Q_{(B)}) N_{2}$.
Then, it can be readily find that $q_{(i)}$ imply the following:
\ben \label{q1}
q_{(t)} &=& 2 + {1 \over \rho^2~ e^{K_{1} + K_{2}}}
\bigg[ \bigg( \eta_{1} - \eta_{2} \bigg)^2 +
\rho^2~ \bigg( e^{K_{1}} - e^{K_{2}} \bigg)^2 \bigg], \\ \label{q2}
q_{(\phi)} &=& 2 + {1 \over e^{L_{1} + L_{2}}}
\bigg[ \bigg( \chi_{1} - \chi_{2} \bigg)^2 +
\bigg( e^{L_{1}} - e^{L_{2}} \bigg)^2 \bigg].
\een
By virtue of the Stoke's theorem one can readily integrate the relation (\ref{jj}) over the chosen
domain of outer communication $<<\cD>>$. Consequently, one arrives at
\ben \label{bou}
\int_{\p <<\cD>>} \na_m q_{(i)}~dS^m &=&   
\int_{\p <<\cD>>}~d\la~\sqrt{{h_{\la \la} \over h_{\mu \mu}}}~\p_\mu q_{(i)} \mid_{\mu = const}
+ \int_{\p <<\cD>>}~d\mu~\sqrt{{h_{\mu \mu} \over h_{\la \la}}}~\p_\la q_{(i)} \mid_{\la = const}
\\ \nonumber
&=&
\int_{<<\cD>>} Tr \bigg( \cJ_{(i)}^{\dagger} \cJ_{(i)} \bigg)~dV.
\een
The main task is to verify if the behaviour of the left-hand side of the above equation on all parts of 
the considered boundary of the two-dimensional manifold $<<\cD>>$. 
In order to perform this task let us rewrite $q_{(i)}$ in the more adequate forms
\ben \label{rel1}
q_{(t)} &=& 
(\Delta \eta)^2~{\rho \over \sqrt{X_1}}{\rho \over \sqrt{X_2}}~e^{\Sigma \varphi}
+ {\sqrt{X_1} \over \rho}~{\rho \over \sqrt{X_2}}~e^{\Delta \varphi}
+ {\sqrt{X_2} \over \rho}~{\rho \over \sqrt{X_1}}~e^{- \Delta \varphi},\\ \label{rel2}
q_{(\phi)} &=& 
(\Delta \chi)^2~{\sqrt{X_1} \over \rho}{\sqrt{X_2} \over \rho}~e^{- \Sigma \varphi}
+ {\sqrt{X_2} \over \rho}~{\rho \over \sqrt{X_1}}~e^{\Delta \varphi}
+ {\sqrt{X_1} \over \rho}~{\rho \over \sqrt{X_2}}~e^{- \Delta \varphi},
\een
where we put for $\Delta f = f_2 - f_1$ and $\Sigma f = f_1 + f_2$.\\
On the black hole event horizon, $X,~A_{(t)},~A_{(\phi)}$ and $\varphi$ are well-behaved functions with
$\cO(1)$-asymptotic. Moreover, $\rho \simeq \cO( \sqrt{\la -c})$ as $\la \rightarrow c$. The same type of behavior one
has for $\sqrt{h_{\mu \mu}/h_{\la \la}} \simeq \cO( \sqrt{\la -c})$. Then, one can conclude that
$\na_m q_{(i)}$ vanishes on the black hole event horizon.
\par
On the symmetry axis, when $\mu \pm 1$, all the quantities under consideration are $\cO(1)$ and
$\rho$ tends to $\cO( \sqrt{1 - \mu})$, as $\mu \rightarrow 1$. On the other hand, when $\mu \rightarrow - 1$,
we have that $\rho \simeq \cO( \sqrt{1 + \mu})$. The square root has the form
$\sqrt{h_{\la \la}/h_{\mu \mu}} \simeq \cO( \sqrt{1 + \mu})$, when $\mu \rightarrow -1$ and
$\sqrt{h_{\la \la}/h_{\mu \mu}} \simeq \cO( \sqrt{1 - \mu})$, when $\mu \rightarrow 1$. Just, having in mind the 
relations (\ref{rel1})-(\ref{rel2}) enables us to find that $q_{(i)}$ remain finite on the axis, implying
that $\na_m q_{(i)} = 0$, for $\mu \pm 1$.
\par
It remains to take into account the contribution for the integration along  $\p \cD^{(4)}$, when
$\la = R \rightarrow \infty$. In the 
aforementioned limit one obtains the following relations for $q_{i}$:
\ben \label{qu1}
q_{(t)} &\simeq& {(\Delta \eta\mid_{\infty})^2 \over 4}~B_{1}^2 B_{2}^2~\la^2~(1-\mu^2)~(1 + \cO(\la^{-1}))
+ \bigg( 
{B_2^2 \over B_1^2} + {B_1^2 \over B_2^2}\bigg)~\cO(1),\\ \label{qu2}
q_{(\phi)} &\simeq& { (\Delta \chi\mid_{\infty})^2 \over \la^2 (1-\mu^2)~(B_1^2 + B_2^2)} ~\bigg(
1 - {2 \over (B_1^2 + B_2^2)\la^2 (1-\mu^2)} \bigg) + \bigg( {B_1^2 \over B_2^2} + {B_2^2 \over B_1^2} \bigg)~\cO(1).
\een
To proceed further, we calculate $\eta$ and $\chi$ and take the limit in question. 
It can be verified that the resulting expressions are provided by
\ben \label{qq1}
\eta \mid_{\la \rightarrow \infty} &\simeq& - {\sqrt{2} \over B^4~(1-\mu^2)}~\mu^2~\cO(\la^{-2}) + 
f(\mu),\\ \label{qq2}
\chi \mid_{\la \rightarrow \infty} &\simeq& - {i \over \sqrt{2}}~B~\bigg( \mu - 
{(1-\mu^2)^{3 \over 2} \over \mu} \bigg)~\la + const,
\een
where $f(\mu)$ is a complicated polynomial in $\mu$-coordinate. It is irrelevant for us because we have partial derivative with respect to $\la$-coordinate, $\p_\la q_{(i)}$, 
in calculation the term for $r \rightarrow \infty$. In view of the above relations (\ref{qu1})-(\ref{qq2}), 
we conclude that $q_{(t)}$ and $q_{(\phi)}$  
tend to a constant value, as $\la \rightarrow \infty$. Then, by virtue of the above, one can deduce that
\be
\int_{<<\cD>>} Tr \bigg( \cJ_{(i)}^{\dagger} \cJ_{(i)} \bigg) = 0,
\label{jj1}
\ee    
which in turn implies that,
$P_{(i) 1} = P_{(i) 2}$ at all points of the region $<<\cD>>$ of the
two-dimensional manifold $\cM$. Just when one considers two black hole solutions 
characterized by $(X_{(1)},~A_{(t)}^{(1)},~A_{(\phi)}^{(1)},~\varphi_{(1)})$
and $(X_{(2)},~A_{(t)}^{(2)},~A_{(\phi)}^{(2)},~\varphi_{(2)})$ of Einstein-Maxwell-dilaton 
equations of motion with a regular domain of outer communication $<<\cD>>$, 
being subject to the same boundary and regularity conditions, they are equal. This envisages
the uniqueness of the dilaton Melvin-Schwarzschild black hole solution.\\
In summary, the direct consequence of our research can be formulated as follows:\\
\noindent
{\bf Theorem}:\\
Let us examine a two-dimensional manifold $\cM$ equipped with a local
coordinate system $(\rho,~ z)$. Suppose further that, the domain of outer communication $<<\cD>>$ is a region in the manifold in question with
boundary $\p \cD$. Let $P_{(i)}$ be hermitian, positive definite
two-dimensional matrices with unit determinants, respectively for 
$i = \tA_t,~ \tA_\phi $ components of Maxwell $U(1)$-gauge field. Moreover, 
let on the boundary of the domain of outer communication
matrices $P_{(i) 1}$ and $P_{(i)2}$ satisfy equation 
$\na_m q_{(i)} = 0$, where $q_{(i)}$ are given by the relations (\ref{q1})-(\ref{q2}).
Then, $P_{(i) 1} = P_{(i) 2}$ in all domain of outer communication  $<<\cD>>$ provided 
that for at least one point $d \in <<\cD>>$
the following is satisfied
$$P_{(i) 1}(d) = P_{(i) 2}(d).$$
Because of the fact that the remaining metric functions are uniquely determined by 
the Ernst's potentials, they have to be identical. This concludes the proof of the uniqueness
of the dilaton Melvin-Schwarzschild black hole. Namely,
all the solutions of dilaton gravity with the same boundary conditions as Melvin universe are the only
static, spherically symmetric black hole in dilaton gravity with non-vanishing $A_t,~A_\phi$
components of $U(1)$ Maxwell gauge field.

\section{Conclusions}
In our paper we have discussed the uniqueness of static axially symmetric black hole spacetime
in dilaton gravity being the low-energy limit of the heterotic string theory. It has been
shown that the underlying equations of motion can be cast in the Ernst's like system of complex relations
which can be written in the form of the matrix equation. Choosing the domain of outer communication 
$<<\cD>>$ as rectangle and using the adequate boundary conditions, we found that
the two solution of the matrix equation, subject to the same boundary and regularity conditions,
are equal in $<<\cD>>$. It enables us to conclude that 
the dilaton Melvin-Schwarzschild black hole which asymptotically tends to dilaton Melvin magnetic universe solution,
is the only axisymmetric static, $S^2$-topology black hole in dilaton gravity with nonzero
$A_0$ and $A_\phi$ components of $U(1)$-gauge field.


\begin{acknowledgments}
 MR was partially supported by the grant of the National Science Center $DEC-2014/15/B/ST2/00089$.\\
 \end{acknowledgments}



\end{document}